\documentclass[9pt]{extarticle}
\usepackage{spconf,amsmath,graphicx}
\usepackage{amssymb}
\usepackage{multirow}
\usepackage{booktabs}
\usepackage{ctable}
\usepackage{stix}
\usepackage{cite}
\usepackage{caption} 
\title{D2Former: A Fully Complex Dual-Path Dual-Decoder Conformer Network using Joint Complex Masking and Complex Spectral Mapping for Monaural Speech Enhancement}
\name{Shengkui Zhao, Bin Ma}
\address{Alibaba Group\\
	\{shengkui.zhao, b.ma\}@alibaba-inc.com\\}
\begin{document}
%
\maketitle
\begin{abstract}
Monaural speech enhancement has been widely studied using real networks in the time-frequency (TF) domain. However, the input and the target are naturally complex-valued in the TF domain, a fully complex network is highly desirable for effectively learning the feature representation and  modelling the sequence in the complex domain. Moreover, phase, an important factor for perceptual quality of speech, has been proved learnable together with magnitude from noisy speech using complex masking or complex spectral mapping. Many recent studies focus on either complex masking or complex spectral mapping, ignoring their performance boundaries. To address above issues, we propose a fully complex dual-path dual-decoder conformer network (D2Former) using joint complex masking and complex spectral mapping for monaural speech enhancement. In D2Former,  we extend the conformer network into the complex domain and form a dual-path complex TF self-attention architecture for effectively modelling the complex-valued TF sequence. We further boost the TF feature representation in the encoder and the decoders using a dual-path learning structure by exploiting complex dilated convolutions on time dependency and complex feedforward sequential memory networks (CFSMN) for frequency recurrence. In addition, we improve the performance boundaries of complex masking and complex spectral mapping by combining the strengths of the two training targets into a joint-learning framework. As a consequence, D2Former takes fully advantages of the complex-valued operations, the dual-path processing, and the joint-training targets. Compared to the previous models, D2Former achieves state-of-the-art results on the VoiceBank+Demand benchmark with the smallest model size of 0.87M parameters.

\end{abstract}   
\begin{keywords}
monaural speech enhancement, conformer, attention, convolution, deep learning
\end{keywords}
\section{Introduction}
\label{sec:intro}

Monaural speech enhancement that aims to extract the signal of interest from noise-corrupted speech is a fundamental and important task. It has been widely deployed in speech communication for better perceptual quality and intelligibility as well as in automatic speech recognition (ASR)  for more robust speech recognition performance. However, monaural speech enhancement is a non-trivial task due to the corruption in both magnitude and phase of speech. 

Early studies only focus on enhancing the magnitude spectrum while reusing the noisy phase spectrum \cite{Huang2014M, Xu2015}.
Recently, Paliwal \textit{et al.} \cite{Paliwal2011K} reports that accurate phase spectrum estimates can considerably contribute towards speech quality. Subsequently, phase enhancement has drawn increasing interest.
Besides the pure phase estimation approaches \cite{Krawczyk2014T, Kulmer2015P},  the time domain approaches \cite{Pascual2017, Rethage2018, Defossez2020, Wang2021B} that directly operate on the raw waveform of speech signals and the time-frequency (TF) domain approaches \cite{Williamson2015, Tan2020D, yin2020phasen, Isik2020, Li2021, Yu2022A, Dang2022H, Cao2022S, Hu2020, Fu2022Y, Zhao2021, Zhao2022B} that manipulate the speech spectrogram are proposed. Although the time-domain approaches have made some success, the TF domain approach has dominated the research trend. It is believed that the fine-detailed structures of speech and noise can be more separable using the short-time Fourier transform (STFT). In order to jointly estimate magnitude and phase spectra, two main categories of targets, complex ratio masking \cite{Williamson2015} and complex spectral mapping \cite{Tan2020D}, have been proposed. Both targets are formulated in the complex domain where the real and imaginary components are to be learnt. The complex ideal ratio mask (cIRM) is shown superior than the real counterpart ideal ratio mask (IRM) \cite{Williamson2015}. On the other hand, the complex spectral mapping is also more effective than the magnitude spectrum mapping \cite{Tan2020D}. However, many recent studies focus on either complex masking or complex spectral mapping, ignoring their performance boundaries. As the training target substantially affects the enhanced results \cite{Wang2014}, developing a training target with higher performance boundary is crucial.
\begin{figure}[t]
  \centering
  \includegraphics[width=0.49\textwidth]{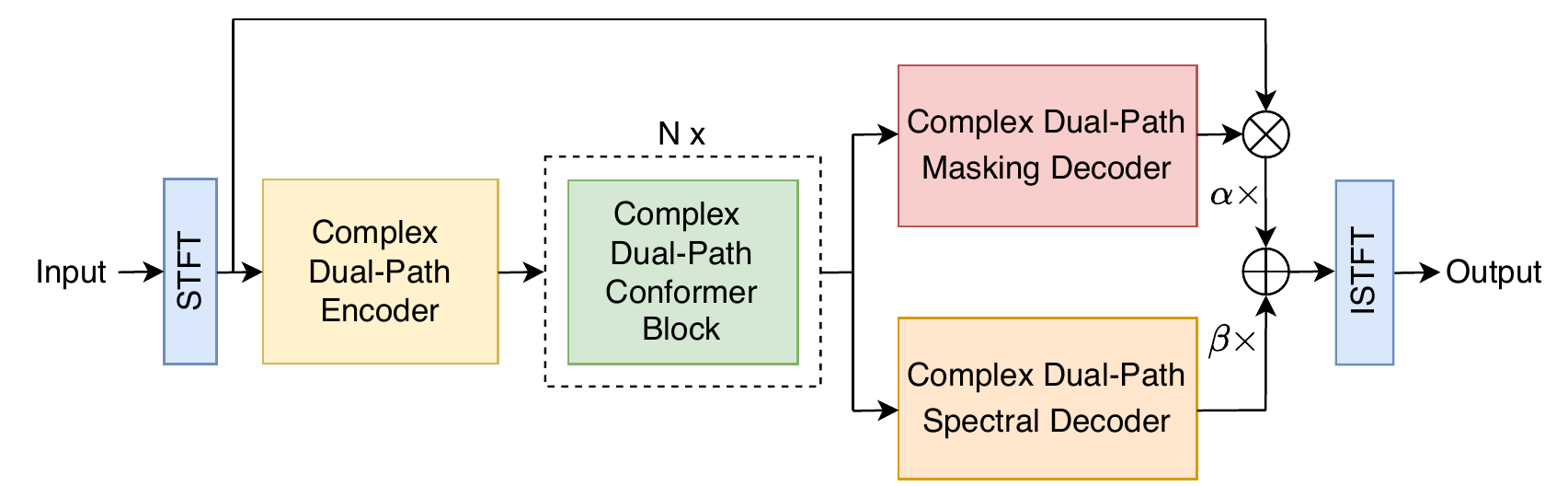}
  \caption{D2Former model architecture. D2Former comprises of a complex dual-path encoder, a complex dual-path conformer module with $N$ repeated blocks, and two complex dual-path decoders. The outputs of the two decoders are weighted and summed.}
  \label{fig1}
\end{figure}

\begin{figure*}[t]
  \centering
  \includegraphics[width=0.8\textwidth]{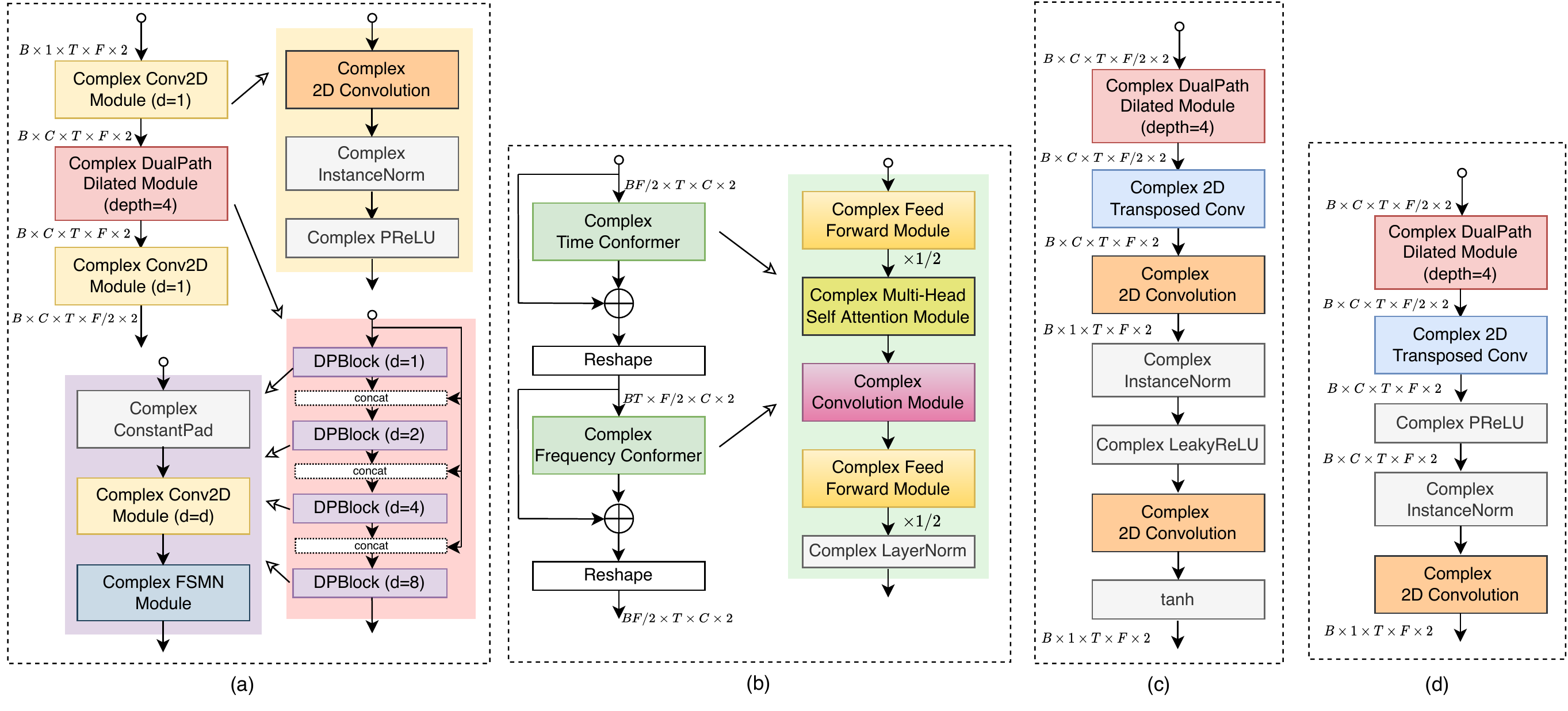}
  \caption{Architecture of D2Former modules. (a) Complex dual-path encoder, (b) Complex dual-path conformer module, (c) Complex dual-path masking decoder, (d) Complex dual-path spectral decoder.}
  \label{fig2}
\end{figure*} 
Typically, most of recent studies treat the real and imaginary parts as two separated real-valued sequences and model them using real-valued networks \cite{Williamson2015, Tan2020D, yin2020phasen, Isik2020, Li2021, Yu2022A, Dang2022H, Cao2022S}. However, the speech spectrogram and the complex targets are naturally complex-valued, much richer representations and more efficiently modelling could be potentially achieved by complex networks \cite{Hu2020, Fu2022Y}. This is because the complex models confine with the complex multiply rules and may jointly learn the real and imaginary parts based on prior knowledge. In our previous studies, we have built complex models based on the convolutional recurrent structure and achieved promising results \cite{Zhao2021, Zhao2022B}.  Recently, the sequence modelling capabilities have been dramatically boosted by Transformer models, especially Conformer \cite{gulati20J}. Unlike the recurrent learning, Conformer employs the self-attention mechanism \cite{Vaswani2017N} to capture global dependency in the sequence while also looking at local dependency using convolutions. Therefore, extending the convolutional recurrent model into a conformer-based model using full complex networks for speech enhancement is highly desirable. 

In this paper, we propose a fully complex dual-path dual-decoder conformer network (D2Former) using joint complex masking and complex spectral mapping for monaural speech enhancement as shown in Fig. \ref{fig1}. In D2Former,  we extend the real-valued conformer network into the complex domain and form a dual-path complex TF self-attention architecture in order to effectively model the complex-valued TF sequence. We further aim to boost the complex encoder and the decoders using a dual-path learning structure by exploiting complex dilated convolutions on time dependency and complex feedforward sequential memory networks (CFSMN) for frequency correlations. In addition, we aim to improve the performance boundaries of complex masking and complex spectral mapping by combining the strengths of the two targets into a joint-learning framework. As a consequence, D2Former takes fully advantages of the complex-valued operations, the dual-path processing, and the joint-training targets. D2Former achieves state-of-the-art (SOTA) results on the VoiceBank+Demand benchmark with the smallest model size of 0.87M parameters.

\section{Proposed D2Former}
The overall D2Former architecture is illustrated in Fig. \ref{fig1}. It consists of a complex dual-path encoder, a complex dual-path conformer block repeated $N$ times, and two separated complex dual-path decoders. The first decoder is for the complex ratio masking and the second decoder is for the complex spectral mapping. 
Specifically, the model takes a noisy speech waveform $y=s+n \in\mathbb{R}^{1\times L}$ and generates an enhanced speech waveform $\hat{s}$. The input waveform is first converted into a complex spectrogram $Y=\mathrm{STFT}(y)\in\mathbb{C}^{T\times F}$, where $T$ and $F$ denote the time and frequency dimensions, respectively. The complex spectrogram sequence $Y$ is then passed into the encoder to extract high-level complex feature sequence. Next, the complex feature sequence is  processed by the complex dual-path conformer blocks.  In each conformer block, the sub-band information in the time path and the full-band information in the frequency path are modelled alternatively. Finally, the complex masks and the complex spectrograms are decoded separately.  The outputs of the dual decoders are weighted and summed to form the final output. 
\begin{figure*}[t]
  \centering
  \includegraphics[width=0.9\textwidth]{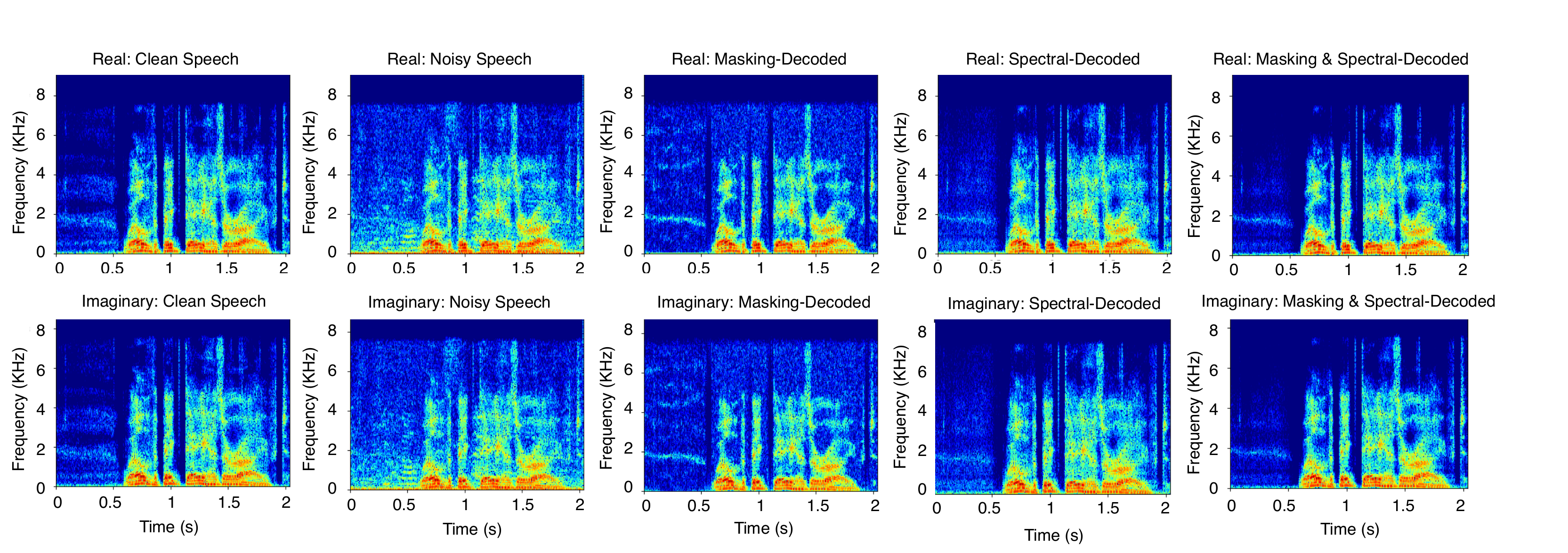}
  \caption{Spectrogram plots of the real (top) and imaginary (bottom) components of clean speech, noisy speech, the complex masking output, the complex spectral mapping output, and the weighted and summed result of the complex masking and complex spectral mapping outputs.}
  \label{fig5}
\end{figure*} 
\subsection{Complex Dual-Path Encoder} 
Our proposed complex dual-path encoder is shown in Fig. \ref{fig2}(a). It consists of two complex 2-dimensional (2D) convolution (Conv2D) modules and a complex dual-path dilated module in between. Each complex Conv2D module consists of a complex  convolution, a complex instance norm, and a complex PReLU. We follow the work \cite{Trabelsi2018} for complex-valued operations. Generally, for a complex  layer $\mathcal{H}$ with trainable parameters, such as the complex convolution layer, and a complex input sequence $Z=Z_R+jZ_I$, the complex operation in $\mathcal{H}$ is represented by
\begin{equation}
\mathcal{H}(Z)=[\mathcal{H}_R(Z_R)-\mathcal{H}_I(Z_I)] + j[\mathcal{H}_R(Z_I)+\mathcal{H}_I(Z_R)]
\label{eqn1}
\end{equation}
where $\mathcal{H}_R$ and $\mathcal{H}_I$ are real and imaginary layers of $\mathcal{H}$ with identical real-valued operations. For a parameterless complex operation $\mathcal{N}$, such as the complex instance norm, the complex operation simply applies $\mathcal{N}$ on the real and imaginary parts separately, that is
\begin{equation}
\mathcal{N}(Z) = \mathcal{N}(Z_R)+j\mathcal{N}(Z_I).
\end{equation}
The first Conv2D module expands the feature map from $1$ channel to $C$ channels. The second Conv2D module halves the frequency dimension from $F$ to $F/2$ for a reasonable complexity. Note that the last dimension of feature map denotes the real and imaginary parts.

The complex dual-path dilated module has a depth of $4$ levels to contain four dual-path blocks (DPBlocks). Each DPBlock contains a complex constant padding, a complex Conv2D module, and a CFSMN module. The complex Conv2D modules respond for extracting feature patterns along the time scale. We increase the receptive field of the Conv2D modules by using dilated convolutions with the dilation factors of \{$1$, $2$, $4$, $8$\}. The CFSMN module contains a CFSMN layer to learn the long-range frequency correlations. The details of the CFSMN layer are provided in our previous 
work \cite{Zhao2022B}. Here we use the same settings for CFSMN. The dual-path design for the encoder not only boosts feature representation along the frequency scale but also expands the time range using dilations.  To aggregate more information and facilitate data flow, the output is concatenated with the input for the top $3$ DPBlocks.
\subsection{Complex Dual-Path Conformer Block}
Conformer was first introduced in \cite{gulati20J} to combine the self-attention mechanism and convolution, which is able to capture both local and global dependencies effectively. The work \cite{Cao2022S} first applies the real-valued conformer to a two-stage modelling scheme. An extended work was proposed in \cite{Fu2022Y} for a semi-complex model. In this work, we develop a fully complex model with a much simpler architecture. Instead of using min-max-norm to compute the complex self-attention \cite{Yang2020M}, we directly apply the softmax function on the absolute values of the complex matrix to closely follow the attention definition and to stabilize the model training \cite{Dong2021Y}. 

Considering that the feature map generated by the encoder is a 3D TF map of dimensions $B\times C\times T \times F/2$ and the self-attention mechanism is a sequence modelling scheme, the dual-path sequence processing \cite{Dang2022H} is more efficient to model such a feature map. Therefore, we apply the dual-path self-attention mechanism on the 3D map to model the time sequence $BF/2\times T\times C$ and the frequency sequence $BT\times F/2\times C$ independently and alternatively. 
Our proposed complex dual-path conformer block is shown in Fig. \ref{fig2}(b). It consists of a complex time conformer and a complex frequency conformer. Each complex conformer extends the original conformer \cite{gulati20J} in the complex domain. Both the feed forward modules and the convolution module are extended into the complex versions based on Eq. \ref{eqn1}. 
 Both the time-sequence attention and the frequency-sequence attention use the same mechanism as described below. Given the complex input $Z=Z_R+jZ_I$ and the complex linear projections $W^Q=W_R^Q+jW_I^Q$,      $W^K=W_R^K+jW_I^K$, and $W^V=W_R^V+jW_I^V$, the queries $Q$, keys $K$ and values $V$ are represented by
\begin{equation}
Q=ZW^Q,\quad K=ZW^K, \quad V=ZW^V.
\end{equation}
Our complex self-attention is computed as
\begin{equation}
\mathrm{ComplexAttention}(Q,K,V)=\mathrm{softmax}\left(\frac{\left|QK^T\right|}{\sqrt{d_k}}\right)V.
\label{eqn4}
\end{equation}
where $|\cdot|$ denotes the absolute value and $d_k$ is the feature dimension. The complex matrix product, $QK^T$, can be calculated as
\begin{equation}
QK^T=(Q_RK_R^T-Q_IK_I^T)+j(Q_RK_I^T+Q_IK_R^T).
\end{equation}
Following the attention design in conformer \cite{gulati20J}, the relative sinusoidal positional encoding scheme is integrated into the real and imaginary parts of $QK^T$, respectively. Following the multi-head self-attention (MHSA) scheme \cite{Vaswani2017N}, the complex MHSA is a concatenation of multiple small-head complex self-attentions. Four heads are used in this work. Moreover, skip connection is used for each conformer to connect the input to the output for ease of training. 
\subsection{Complex Dual-Path Masking and Spectral Decoders}  
The decoder aggregates information from the processed feature map to predict the target. In this work, we propose a dual-decoder scheme using both the complex masking \cite{Williamson2015} and the complex spectral mapping \cite{Tan2020D} to estimate the magnitude and phase spectra. We believe that the dual-decoder scheme not only encourages learning a more general feature representation but also compensating the output for any information loss, thus improving the performance boundary. 

Our proposed masking and spectral decoders are shown in Fig. \ref{fig2} (c) and (d), respectively. Both decoders consist of a complex dual-path dilated module, similar to the one in the encoder, to aggregate information from both time and frequency scales. The complex transposed Conv2D layer is used in both decoders to recover the frequency dimensions. In the masking decoder, a complex Conv2D layer is used to squeeze the channel to $1$, which is then followed by a complex instance norm and a complex LeakyReLU activation. The last complex Conv2D layer projects the output to the masking domain, followed by \textit{tanh} activation to bound the cIRM estimates. For the spectral decoder, we first apply the complex PReLU and the complex instance norm, and then the complex Conv2D layer to squeeze the channel. No activation is used for the output of the spectral decoder. 
\begin{table*}
\center
\scriptsize
\begin{tabular}{lclcccccccccc}
\specialrule{.1em}{.05em}{.05em}
\multirow{2}{*}{Model} &\multirow{2}{*}{Year} &\multirow{2}{*}{Prediction} &\multirow{2}{*}{Para.(M)} &\multirow{2}{*}{Self-Attention} &\multirow{2}{*}{DualP E\&D}&\multirow{2}{*}{$\alpha$} &\multirow{2}{*}{$\beta$} & \multicolumn{5}{c}{Evaluation Metrics}         \\
\cline{9-13}
                                &          &                                 &      &        &    &     &      &WB-PESQ &CSIG &CBAK &COVL &STOI       \\ \hline 
Noisy                           & -        &\quad -                          &-     &-       &-   & -   &-     &1.97 &3.35 &2.44 &2.63 &0.91     \\                                 
SEGAN \cite{Pascual2017}        & 2017     &Waveform                         &97.5  &No       &No  & -   &-     &2.16 &3.48 &2.94 &2.80 &0.92     \\
DEMUCS \cite{Defossez2020}      & 2020     &Waveform                         &128   &No       &No  & -   &-     &3.07 &4.31 &3.40 &3.63 &0.95     \\
TSTNN \cite{Wang2021B}          & 2021     &Waveform                         &0.92  &Real    &No  & -   &-     &2.96 &4.10 &3.77 &3.52 &0.95     \\
GaGNet \cite{Li2021}            & 2021     &Complex Mask                     &5.94  &No       &No  & -   &-     &2.94 &4.26 &3.45 &3.59 &-     \\
FRCRN \cite{Zhao2022B}          & 2022     &Complex Mask                     &6.98  &No       &Yes & -   &-     &3.21 &4.23 &3.64 &3.73 &-     \\
DB-AIAT \cite{Yu2022A}          & 2022 &Magnitude Mask + Complex Spectrum    &2.81  &Real    &No  & -   &-     &3.31 &4.61 &3.75 &3.96 &\textbf{0.96}     \\
DPT-FSNet \cite{Dang2022H}      & 2022 &Complex Mask                         &0.91  &Real    &No  & -   &-     &3.33 &4.58 &3.72 &4.00 &\textbf{0.96}     \\    
CMGAN \cite{Cao2022S}           & 2022 &Magnitude Mask + Complex Spectrum    &1.83  &Real    &No  & -   &-     &3.41 &4.63 &\textbf{3.94} &4.12 &\textbf{0.96}     \\ \hline 
\textbf{D2Former}               & 2022     &Complex Mask + Complex Spectrum  &0.81  &Real    &No  & 0.95&0.05  &3.33 &4.60 &3.81 &3.98 &\textbf{0.96}   \\ \cline{4-13}
                                &          &                                 &0.81  &Complex &No  & 0.95&0.05  &3.37 &4.63 &3.89 &4.12 &\textbf{0.96}     \\  \cline{4-13}
                                &          &                                 &0.87  &Complex &Yes & 0.95&0.05  &3.41 &4.65 &3.91 &4.14 &\textbf{0.96}   \\  
                                &          &                                 &      &        &    & 0.75&0.25  &\textbf{3.43} &\textbf{4.66} &\textbf{3.94} &\textbf{4.15} &\textbf{0.96}   \\  
                                &          &                                 &      &        &    & 0.50&0.50  &3.42 &\textbf{4.66} &3.92&4.14&\textbf{0.96}     \\   
                                &          &                                 &      &        &    & 0.25&0.75  &3.42 &4.65 &3.91 &4.14&\textbf{0.96}     \\   
                                &          &                                 &      &        &    & 0.05&0.95  &3.41 &4.63 &3.92 &4.13&\textbf{0.96}     \\  \hline 
\specialrule{.1em}{.05em}{.05em}
\label{tab2}
\end{tabular}
\caption{Performance comparison on the VoiceBank+Demand benchmark dataset. Prediction denotes the model output before STFT; DualP E\&D denotes dual-path encoder-decoder. Here D2Former uses a different DualP E\&D as FRCRN; The $\alpha$ and $\beta$ only apply to D2Former.}
\end{table*}

Let the output of the masking decoder be denoted by $M=M_R+jM_I$. By applying the mask $M$ on the noisy spectrogram $Y$, we obtain the enhanced spectrogram $\hat{S}^\prime=\hat{S}_R^\prime+j\hat{S}_I^\prime=M\odot Y$, where $\odot$ is element-wise complex multiplication. Let the output of the spectral decoder be denoted by $\hat{S}^\dprime=\hat{S}_R^\dprime+j\hat{S}_I^\dprime$. By element-wise summing the two weighted outputs $\hat{S}^\prime$ and $\hat{S}^\dprime$, we obtain the final enhanced complex spectrogram:
\begin{equation}
\hat{S}=\alpha\hat{S}^\prime + \beta\hat{S}^\dprime = (\alpha\hat{S}_R^\prime+\beta\hat{S}_R^\dprime)+j(\alpha\hat{S}_I^\prime+\beta\hat{S}_I^\dprime)
\end{equation}
where $\alpha$ and $\beta$ are weighting factors.
\section{Experiments}
We conduct evaluation and comparison of our proposed D2Former with the previous models on the popular speech enhancement benchmark: VoiceBank+Demand dataset \cite{Botinhao2016}. The baseline models with the published results on the VoiceBank+Demand dataset are selected. The dataset consists of 11,472 clean-noisy matched training mixtures from 28 independent speakers and 10 types of noise. The mixtures are created at SNRs from 0 dB to 15 dB with 5 dB incremental steps. The testset consists of 824  mixtures from 2 unseen speakers and 5 unseen noise types. The SNRs span from 2.5 dB to 17.5 dB with 5 dB incremental steps. All utterances are resampled to 16 kHz and 2s long segments are randomly sampled during training.

We use a Hamming window with length of 25 ms and hop size of 6.25 ms. The STFT length is set to 400 points, resulting $F=201$ and $F/2=101$. The hyper-parameter settings are: $B=2$, $N=3$, and $C=32$. Our model is trained for maximum of 120 epochs using the AdamW optimizer with an initial learning rate of $5e^{-4}$. The learning rate holds for the beginning 30 epochs and is then decayed by a factor of 0.5 with patience of 1 epoch.  Our enhanced audio samples are available online \footnote{https://github.com/alibabasglab/D2Former}.
\subsection{Training Loss}
Inspired by Cao \cite{Cao2022S}, we use a combination of the time domain loss $\mathcal{L}_\mathrm{Time}$, the TF domain loss $\mathcal{L}_\mathrm{TF}$, and the quality net (QNet) loss $\mathcal{L}_\mathrm{QNet}$:
\begin{align}
&\mathcal{L}=\mathcal{L}_\mathrm{TF} + \gamma_1\mathcal{L}_\mathrm{Time} + \gamma_2\mathcal{L}_\mathrm{QNet} \nonumber \\ 
&\mathcal{L}_\mathrm{TF}=\mathrm{MSE}\left(|S|^P, |\hat{S}|^P\right) + \gamma_3\left[\mathrm{MSE}\left(S_R, \hat{S}_R\right)+\mathrm{MSE}\left(S_I, \hat{S}_I\right)\right]\nonumber \\ 
&\mathcal{L}_\mathrm{Time}=\mathrm{Mean}\left(\left|s-\hat{s}\right|\right) \nonumber \\ 
&{L}_\mathrm{QNet}=\mathrm{MSE}\left(\mathrm{QNet}\left(\mathrm{Concat}\left(|S|^P, |\hat{S}|^P\right) \right)-1\right) 
\end{align}
where $|\cdot|$ denotes magnitude, and $P=0.3$ is magnitude compression factor. The following weighting factors $\gamma_1=0.2$, $\gamma_2=0.05$ and $\gamma_3=0.1$ are used for balancing the loss contributions. We use the same network and the same training method for QNet as in \cite{Cao2022S}. 
\subsection{Results \& Ablation Studies}  
Five metrics are utilized, namely WB-PESQ, CSIG, CBAK, COVL, and STOI  \cite{Cao2022S}.  Table 1 shows the performance comparison with the published results of the SOTA models. We categorize the models based on the prediction, the applied self-attention, and the applied dual-path scheme in encoder-decoder (DualP E\&D). Generally speaker, the models that predict the magnitude or complex TF masks show better performance than the models that predict the waveform directly, except for GaGNet \cite{Li2021}. The models that are built on self-attentions not only show improved performance but also have much smaller model sizes compared to the models that do not use self-attentions. Note that most of the previous models are based on real-valued self-attentions. To show the effectiveness of the complex self-attention, we list the results of the D2Former with the real-valued self-attention where the queries $Q$, keys $K$ and values $V$ use absolute values. One can find that D2Former with the complex attention improves the evaluation metrics. For example, WB-PESQ increases from 3.33 to 3.37. To show the effectiveness of DualP E\&D, we list the results of D2Former without DualP E\&D where the CFSMN module is removed. It shows that DualP E\&D helps improve the evaluation metrics with marginal parameters added. 

Finally, we evaluate the affects of the weighting factors $\alpha$ and $\beta$. It is known that a large $\alpha$ weights more on the complex masking decoder and vice versa. The results show that neither the individual complex masking decoder nor the individual complex spectral decoder works the best. The weighted results with $\alpha=0.75$ and $\beta=0.25$ achieve the best with one selected example shown in Fig. \ref{fig5}. The results imply that a better trade-off can be achieved with a joint-learning framework compared to an individual learning target. Note that DB-AIAT \cite{Yu2022A} and CMGAN \cite{Cao2022S} also use a joint-learning framework. However, they use magnitude mask instead of complex mask. From the point view of model parameters, D2Former uses the least number of parameters of 0.87M. Compared to CMGAN \cite{Cao2022S}, D2Former uses 48\% parameters but achieves better metric results.    
  
\section{Conclusions}
In this work, we introduced a fully complex D2Former, built on conformer network for monaural speech enhancement. Contrary to prior models, we learn the complex masking and complex spectral mapping in a single model architecture. By leveraging the dual-path complex self-attention,  the dual-path encoder and decoder, and the joint-learning framework for complex masking and complex spectral mapping, we successfully enhanced the model capability and  achieved  SOTA results on the VoiceBank+Demand benchmark with the least number of model parameters.
\bibliographystyle{IEEEbib}
\bibliography{refs}

\end{document}